\documentclass[aps,amsmath,amssymb,showpacs,twocolumn]{revtex4}

\usepackage{graphicx}
\usepackage[latin1]{inputenc}

\newcommand{\ie}{\textsl{i.e.}}

\newcommand{\eg}{\textsl{e.g.}}
\newcommand{\fref}{Fig.}
\newcommand{\eref}{Eq.}

\begin{document}


\title{Laser spectroscopy of gas confined in nanoporous materials} 


\author{Tomas Svensson}
\email[E-mail: ]{tomas.svensson@fysik.lth.se}%
\homepage[Webpage: ]{www.atomic.physics.lu.se/biophotonics}%
\affiliation{Division of Atomic Physics, Department of Physics, Lund University, Sweden}%
\author{Zhijian Shen}%
\affiliation{Department of Physical, Inorganic and Structural Chemistry, Arrhenius Laboratory, Stockholm University, Sweden.}%
\affiliation{Berzelii center EXSELENT on porous materials, Stockholm University, Sweden}

\date{\today}

\begin{abstract}
We show that high-resolution laser spectroscopy can probe surface interactions of gas confined in
nano-cavities of porous materials. We report on strong line broadening and unfamiliar lineshapes
due to tight confinement, as well as signal enhancement due to multiple photon scattering. This new
domain of laser spectroscopy constitute a challenge for the theory of collisions and spectroscopic
lineshapes, and open for new ways of analyzing porous materials and processes taking place therein.
\end{abstract}



\maketitle

The tight confinement and large surface-areas offered by nanoporous materials give rise to numerous
interesting and useful phenomena. Important topics in this vast and rapidly advancing field include
storage of carbon dioxide and hydrogen \cite{Rosi2003_Science,Wang2008_Nature}, molecular sieving
\cite{Davis2002_Nature}, catalysis \cite{Corma1997_ChemRev,Thomas1999_Nature}, ion exchange,
melting and freezing under confinement \cite{Christenson2001_JPhysCondensMat}, luminescence of
porous silicon \cite{Canham1990_ApplPhysLett}, and chemical sensing. The interactions between gases
and porous materials are often of particular interest. However, due to lack of methodologies, many
processes occurring inside nanoporous structures are not fully understood.

Here, for the first time, we show that the high-resolution laser absorption spectroscopy allows
sensing of gases confined in nano-cavities of porous materials and investigation of the gas-surface
interactions. We report that the tight confinement, \ie\ frequent surface interactions, results in
strong (GHz) line broadening and unfamiliar lineshapes. Severe multiple scattering, and the
corresponding long photon pathlengths, enhance our signals and makes it possible to study weak
transitions and thin samples. The interpretation of observed high-resolution spectra constitutes a
new challenge for the theory of collisions and spectroscopic lineshapes. This completely unexplored
domain of laser spectroscopy envisages new ways for studying important processes occurring in
nanoporous materials, such as gas adsorption, gas transport, heterogenous catalysis, and van der
Waal interactions between surfaces and molecules. In addition, nanoporous materials may be an
interesting system for fundamental studies of \eg\ hard collisions and the corresponding effects on
molecular spectra. Finally, as the line broadening is shown to depend heavily on the pore size, our
work reveals a new, non-destructive technique for pore size assessment.

High-resolution laser absorption spectroscopy is already a well established tool for sensitive,
selective and rapid sensing of gases and atomic vapors, as well as for investigations of
intermolecular collision processes. Conventionally, experiments involve open path monitoring or
measurements on free gases or vapors contained in single- or multipass gas cells. More recently, it
has also been demonstrated that the technique can be used also for sensing of gases dispersed
inside turbid solid materials \cite{Sjoholm2001_OptLett}. The contrast between the sharp absorption
features of the gas phase and the slowly varying spectra of solids renders detection of minute gas
absorption feasible, even if the solid exhibits a much stronger absorption. Instead, the main
difficulty in such experiments is the severe optical interference effects (speckle) originating
from multiple scattering in the porous material \cite{Svensson2008_OptLett}.

In this Letter, we present investigations of molecular oxygen (O$_2$) located in subwavelength
pores of nanoporous bulk alumina (Al$_2$O$_3$). Measurements have been conducted on two materials
with different pore size distribution, both carefully characterized using mercury intrusion
porosimetry \cite{Rouquerol1994_PureApplChem}. The first is a $\alpha$-alumina ceramic manufactured
by sintering a monodispersive 0.3 $\mu$m $\alpha$-alumina powder at 1000 $^\circ$C. The resulting
cylindrical samples are 13 mm in diameter with a thickness up to 6 mm, having a narrow ($\pm 10$
nm) pore size distribution centered around 70 nm and a total porosity of 35\%. The second material
is a $\gamma$-alumina ceramic prepared by means of spark plasma sintering, yielding a 2.2 mm thick,
12 mm diameter sample with pore sizes of $18\pm2$ nm and a total porosity of 50\%.

In order to communicate with oxygen confined in these materials, we utilize the R9Q10 line (760.654
nm) of the near infrared A-band of O$_2$, which originate from a $0\leftarrow 0$ vibrational
transition of the $b^1\Sigma_g^+\leftarrow X^3\Sigma_g^-$ electronic transition. The transition is
electric dipole forbidden and weak, having a linestrength of about $2.6\times10^{-13}$ cm$^2$Hz at
room temperature \cite{Predoi-Cross2008_JMolSpectrosc}. At atmospheric conditions (1 atm, 300 K,
20.9\% O$_2$), this results in a peak absorption coefficient of $2.7\times 10^{-4}$ cm$^{-1}$ and a
half width at half maximum (HWHM, $\Gamma$) of 1.6 GHz (a 0.4 GHz Doppler component together with a
1.5 GHz pressure broadening, yielding a Voigt profile that can be approximated by a Lorentz
profile). Our spectroscopic instrumentation is based on high-resolution tunable diode laser
absorption spectroscopy (TDLAS), and is combined with wavelength modulation spectroscopy (WMS) when
higher sensitivity is needed. A detailed description of the instrumentation and procedures for WMS
signal processing is available elsewhere \cite{Svensson2008_APB}. Briefly, the output from a 0.3 mW
tunable single-mode vertical-cavity surface-emitting diode laser (VCSEL) is injected centrally into
the porous solid, and the diffuse transmission is detected by a 5.6$\times$5.6 mm$^2$ large-area
photodiode. All measurements are carried out under atmospheric conditions, and WMS experiments are
conducted using a optical frequency modulation amplitude of 2.2$\times$1.6 GHz (\ie\ maximizing
then second harmonic WMS signal of free oxygen). Speckle noise is suppressed by employing
mechanical dithering and averaging \cite{Svensson2008_OptLett}.

\begin{figure}
  \includegraphics[]{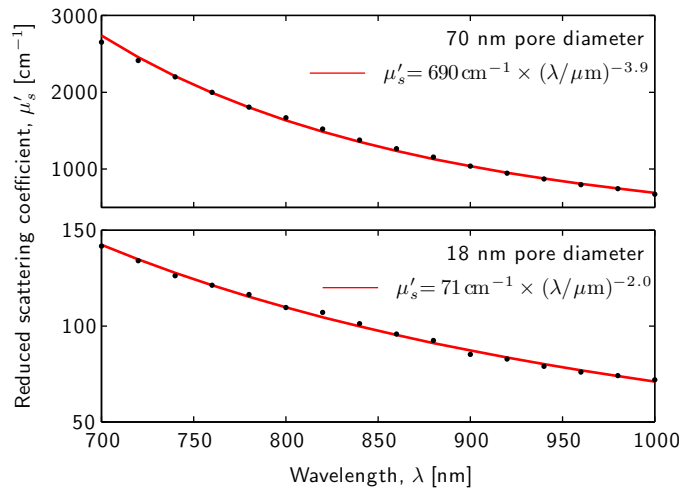}
  \caption{Reduced scattering coefficients of the two alumina materials (pore diameters of 70 nm and 18 nm, respectively).
  At 760 nm, the difference in scattering
  is a factor 16 (2000 cm$^{-1}$ against 120 cm$^{-1}$). Interestingly, the sample with the larger pores exhibit a $\lambda^{-4}$
  Rayliegh-type scattering decay, while the sample with the smaller pores only show a $\lambda^{-2}$
  behaviour. This suggests that the major scattering contribution in the latter sample originates from collective
  effects rather than from individual 18 nm pores.\label{fig:opt_prop}}
\end{figure}

Since the strength of the gas absorption imprint is proportional to the pathlength through pores,
the scattering properties of the samples are of paramount importance. We have measured the optical
properties of our samples by employing photon time-of-flight spectroscopy (PTOFS) based on a pulsed
supercontinuum fiber laser, acousto optical tunable filtering and time-correlated single photon
counting \cite{Svensson2009_RevSciInstrum}. We inject short (ps) light pulses centrally into the
cylindrical samples using a 600 $\mu$m optical fiber, and collect the transmitted and broadened
pulses on the other side using a second 600 $\mu$m fiber. The measurement gives us direct
information on the time-of-flight (TOF) distribution. Using a volume averaged refractive index,
$\phi\times n+(1-\phi)\times 1$ where $n$ is the refractive index of the solid and $\phi$ the
porosity, we can estimate the corresponding pathlengths. Via diffusion theory of light transport,
we can also determine reduced scattering coefficients. \fref\ \ref{fig:opt_prop} shows the
scattering properties of our two alumina materials.

\begin{figure}
  \includegraphics[]{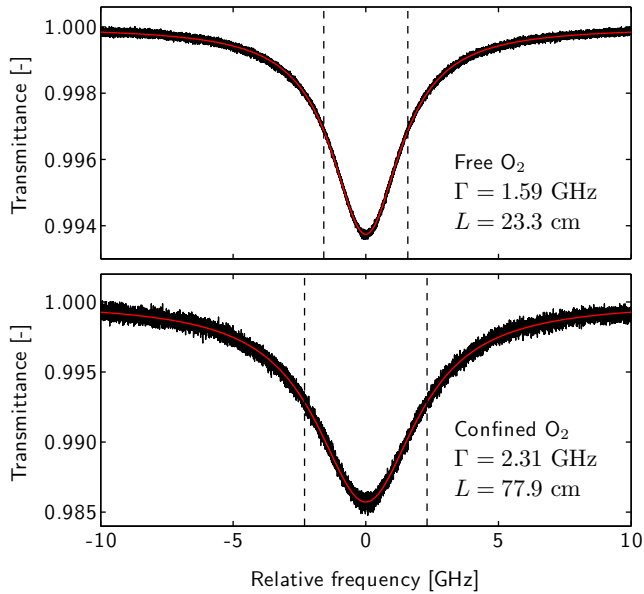}
  \caption{Comparison of experimental lineshapes of free oxygen and oxygen confined in 70 nm pores of
  $\alpha$-alumina. Lorentzian profiles (red) are fitted to experimental data (black). Fitted lineshape halfwidths,
  $\Gamma$, are stated in the graph and marked by vertical dashed lines. The fitted pathlength through pores, $L$, assumes
  an oxygen concentration of 20.9\%. For the $\alpha$-alumina sample, the detected power was 0.8 $\mu$W.\label{fig:direct_signals}}
\end{figure}

The $\alpha$-alumina exhibit extremely strong scattering, and in order to fit the TOF distribution
within the 12.5 ns window offered by our 80 MHz rep. rate laser we could not perform PTOFS
measurements on samples more than 2 mm thick. The reduced scattering coefficient at 760 nm is about
2000 cm$^{-1}$, meaning that the light propagation can be though of as a random walk of photons
being isotropically scattering on average each 5 $\mu$m. For a 2 mm $\alpha$-alumina sample, the
average TOF was about 1.4 ns, corresponding to a pathlength of 28 cm. Since average pathlength is
proportional to the square of the thickness, the average pathlength of photons transmitted to a 6
mm thick sample can be expected to exceed 200 cm. Such long pathlengths allows us to study the
direct oxygen absorption. \fref\ \ref{fig:direct_signals} shows the experimental high-resolution
spectrum of O$_2$ confined in the 70 nm diameter pores of a 6 mm thick $\alpha$-sample, and
compares it to a measurement on free oxygen (a ~23 cm path through ambient air). Data evaluation
involves simultaneous fitting of a linear baseline and a superimposed Lorentzian lineshape. While
experiments on free oxygen are in good agreement with expected line parameters, oxygen confined in
the $\alpha$-alumina exhibit a significantly broader spectrum (2.3 GHz HWHM, compared to 1.6 GHz
for free oxygen). Although mercury intrusion porosimetry already have shown that the material has
an open porosity, we double checked this issue by flushing the sample with nitrogen. The oxygen
imprint vanished completely almost instantaneously, further proving that we are not measuring
oxygen in closed pores with elevated pressures.

\begin{figure}
  \includegraphics[]{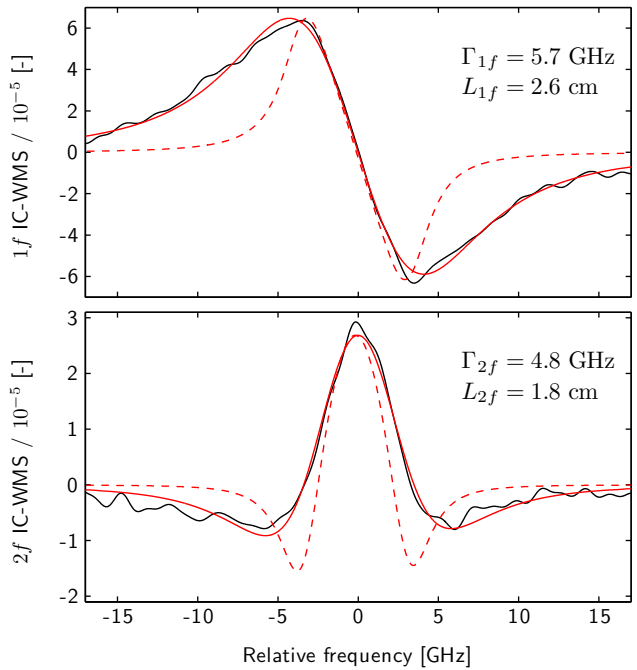}
  \caption{Intensity-corrected WMS signals (IC-WMS) obtained from the $\gamma$-alumina sample with 18 nm pore diameters
  (black solid) together with fitted WMS simulations for Lorentzian lineshapes (red solid). The detected power was 3 $\mu$W. The tight confinement causes
  significant line broadening, here illustrated by a comparison to normalized WMS signals of free oxygen (dashed
  red, $\Gamma=1.6$ GHz). Stated HWHM ($\Gamma$) and pathlength through pores ($L$) refer to fitted values. Note the discrepancy
  in fitted values between first and second harmonic, showing that a Lorentzian is an improper model
  of the lineshape.
  \label{fig:WMS_data}}
\end{figure}

The lower scattering of the $\gamma$-alumina material makes gas sensing more challenging. The
average TOF at 760 nm was 120 ps, corresponding to a total pathlength of about 3 cm. Although the
large-area photodiode collects light with slightly longer average pathlengths, the pathlength
through pores is not expected to exceed 2 cm. This corresponds to a peak absorption fraction on the
order of $10^{-4}$ or less. To be able to study such a weak absorption, we were forced to employ
wavelength modulation spectroscopy (WMS). While increasing sensitivity, WMS renders lineshape
analysis less straightforward. Furthermore, in order to eliminate non-negligible oxygen absorption
originating from optical path outside the sample, WMS measurements were made in a differential
manner (the pathlength offset is determining by measuring on a non-porous solid of equal
thickness). \fref\ \ref{fig:WMS_data} shows the resulting first ($1f$) and second ($2f$) WMS
harmonics together with fitted WMS simulations. The tight confinement is now the dominating source
of line broadening. The linewidth is about 5 GHz, by far exceeding the 1.6 GHz exhibited by free
oxygen at atmospheric pressure. Although Lorentzian lineshapes appear to model the WMS data fairly
well (at least in $1f$), the major discrepancy in fitted HWHM ($\Gamma$) and pathlength ($L$)
clearly shows that a Lorentzian is an improper lineshape model. In contrast, similar WMS
experiments on the $\alpha$-alumina (where the relative Lorentzian contribution from pressure
broadening is strong) results in minor differences between $1f$ and $2f$ (less than 2\% and 7\%
difference in $\Gamma$ and $L$, respectively), and a HWHM in good agreement with the direct
transmittance shown in \fref\ \ref{fig:direct_signals}.

As intermolecular collisions is known to be a strong source of line broadening, it is natural to
investigate whether wall collisions may explain the broadening exhibited by these tightly confined
oxygen molecules. In fact, the mean free path of O$_2$ in air at atmospheric pressure is about 60
nm. This means that nano-confinement clearly may render wall collisions and intramolecular
collisions comparable sources of broadening. Broadening of spectral lines due to wall collisions
have previously been discussed for the case of low pressure microwave spectroscopy. There,
investigations in the mTorr regime (10$^{-6}$ atmospheres) renders the mean free path comparable to
the size of the utilized gas cells (\ie\ cm scale). The exact nature of the broadening mechanism
has been the subject of a lively discussion
\cite{Gordy1948_RevModPhys,Johnson1952_PhysRev,Danos1953_PhysRev}. According to classical gas
kinetics, the rate $f_\textrm{wall}$ at which molecules in an ideal gas collides with the walls of
its container follows
\begin{equation}\label{eq:f_wall}
    f_\textrm{wall} = \frac{1}{\tau_\textrm{wall}} = \frac{A}{V}\sqrt{\frac{kT}{2\pi m}}
    =\frac{A}{4V}\times v_\textrm{avg}
\end{equation}
where $\tau_\textrm{wall}$ is the average time between wall collisions, $A$ the container area, $V$
the volume, $k$ the Boltzmann constant, $T$ the temperature, $m$ the molecular mass, and
$v_\textrm{avg}$ the average speed of O$_2$ molecules (450 nm/ns at 300 K). It was first "tacitly"
assumed that wall collision gave rise to a Lorentzian component with a HWHM $\Delta\nu$ simply
given by $1/2\pi\tau_\textrm{wall}$, and that the combined collisional component could be described
by Lorentzian with
$\Delta\nu_\textrm{coll}=(\Delta\nu_\textrm{wall}^2+\Delta\nu_\textrm{mol}^2)^\frac{1}{2}$. It was
later theoretically shown that the lineshape is not Lorentzian \cite{Danos1953_PhysRev}, and that
the collisional components rather are additive,
$\Delta\nu_\textrm{coll}\simeq\Delta\nu_\textrm{wall}+\Delta\nu_\textrm{mol}$
 (the exact relation depending both on the container shape and the
relative strength of the two contributions) \cite{Luijendijk1975_JPhysB}. The combined effects of
wall and intermolecular collisions could, however, in practice always be approximated by the
Lorentzian lineshape.

In contrast to microwave spectroscopy, where reported levels of wall broadening were in the kHz
range, our experiments show that molecules confined in nanopores exhibit broadening in the GHz
range, \ie\ a factor $10^6$ stronger.  A theoretical estimation is given by employing
$\Delta\nu=1/2\pi\tau$ in combination with the outcome of \eref\ \ref{eq:f_wall} for \eg\ spherical
cavities. For 70 nm and 18 nm sphere diameters, the procedure yields $\Delta\nu$'s of 1.5 and 6
GHz, respectively. However, since the pore size distribution given by mercury intrusion porosimetry
measures the most narrow passages in the pore structure \cite{Rouquerol1994_PureApplChem}, and
since the actual pore shape is unknown, it can not be expected that our experiment should agree
with the $\Delta\nu$ expected for 70 nm and 18 nm spheres, respectively. Instead, we expect the
effective pore size to be larger, and wall broadening therefore less pronounced. Although the exact
strength of the broadening cannot be derived from our experiment (since the lineshape is unknown
and added on top of pressure broadening), these basic estimations show that the observed levels of
broadening is in agreement with theory of wall collision broadening. On the other hand, we have not
yet been able to explain the discrepancy between WMS harmonics. This means that further work on
lineshape models for confined molecules are needed.

As current theory of collisions and their impact on lineshape is not aimed at explaining the
frequent surface interactions of tightly confined molecules, we anticipate that continued research
on high-resolution spectroscopy of tightly confined gas will offer new surprises. It should be
noted that all measurements described above was conducted at atmospheric pressure. In contrast to
intramolecular broadening, broadening due to wall interactions is pressure independent. It is thus
possible to further isolate the phenomenon by performing experiments under reduced pressures. This
will allow more careful analysis of lineshapes, as well as investigations of surface interactions
in larger pores of macroporous materials. In fact, broadening due to wall collisions in
micrometer-sized cavities was recently encountered in connection to low pressure spectroscopy of
gas confined in the 10 $\mu$m core of photonic crystal fibers
\cite{Ghosh2005_PhysRevLett,Hald2007_PhysRevLett}. Besides being of fundamental physical interest,
broadening due to wall interactions can also be used for non-destructive assessment of effective
pore size. The use of low pressures makes this approach feasible even for materials with larger
pores. Furthermore, although no line shift was observed in our experiments, van der Waals
interactions between surfaces and atoms or molecules may be an important effect if our experiments
are extended to the study of other gases or vapors. Line shifts due to coupling between atomic
vapors and solid surfaces has been demonstrated for atomic beams passing close to surfaces
\cite{Sandoghdar1992_PhysRevLett}, as well as for vapors confined in so called "extremely thin
cells" \cite{Maurin2005_JPhys,Fichet2007_EurophysLett}. Similar experiments made on gases confined
in porous materials may be of great value for the understanding of important processes such as gas
adsorption. Another aspect not yet mentioned is Dicke narrowing \cite{Dicke1953_PhysRev}, which has
been shown to be influenced by the interaction with surfaces and atoms
\cite{Frueholz1987_PhysRevA}. Its relation to our experiments on gases in porous materials remains
an open question.

This work was funded by the Swedish Research Council via Prof. Stefan Andersson-Engels. Dr. Mats
Andersson, Dr. Lars Rippe, MSc. Erik Alerstam, Dr. Dmitry Khoptyar, Prof. Stefan Kr\"{o}ll, Prof.
Sune Svanberg, and Prof. Stefan Andersson-Engels are acknowledged for fruitful discussions and/or
collaboration on spectroscopic instrumentation. Karin Lindqvist at Swerea IVF is acknowledged for
preparing the $\alpha$-alumina.


\end{document}